\documentclass[a4paper,fleqn]{cas-sc}

\usepackage[numbers]{natbib}
\usepackage{stfloats}

\newcommand*\p{{(+)}}
\newcommand*\m{{(-)}}
\newcommand*\both{{(\pm)}}
\newcommand*\dr{\mathrm{(drift)}}
\newcommand*\lc{\mathrm{(lock)}}

\begin{document}
\let\WriteBookmarks\relax
\def\floatpagepagefraction{1}
\def\textpagefraction{.001}

\shorttitle{Predicting the Oscillatory Regimes of Global Synchrony Induced by Secondary Clusters} 

\shortauthors{G.Y. Kim, M.J. Lee, and S.-W. Son}  

\title [mode = title]{Predicting the Oscillatory Regimes of Global Synchrony Induced by Secondary Clusters}



%

\author[1]{Gug Young Kim}





\credit{Conceptualization, Data curation, Formal analysis, Investigation, Methodology, Project administration, Validation, Visualization, Writing – original draft}
 
\author[2]{Mi Jin Lee}[orcid=0000-0003-0013-7907]

\cormark[1]




\credit{Conceptualization, Formal analysis, Investigation, Methodology, Project administration, Supervision, Validation, Writing – review \& editing, Funding acquisition}

\author[1, 3, 4]{Seung-Woo Son}[orcid=0000-0003-2244-0376]

\cormark[2]




\credit{Conceptualization, Formal analysis, Investigation, Methodology, Project administration, Supervision, Validation, Writing – review \& editing, Funding acquisition}

\affiliation[1]{organization={Department of Applied Physics, Hanyang University},
            city={Ansan},
            postcode={15588}, 
            country={Korea}}
\affiliation[2]{organization={Department of Physics, Pusan National University},
            city={Busan},
            postcode={46241}, 
            country={Korea}} 
\affiliation[3]{organization={Division of Defense Intelligence and Information Convergence Engineering, Hanyang University},
            city={Ansan},
            postcode={15588}, 
            country={Korea}}
          
\affiliation[4]{organization={Asia Pacific Center for Theoretical Physics},
            city={Pohang},
            postcode={37673}, 
            country={Korea}}    
\cortext[1]{mijinlee@pusan.ac.kr}
\cortext[2]{sonswoo@hanyang.ac.kr}



\begin{abstract}
Synchronization systems with effective inertia, such as power grid networks and coupled electromechanical oscillators, are commonly modeled by the second-order Kuramoto model. In the forward process, numerical simulations exhibit a staircase-like growth of global synchrony, reflecting temporal oscillations induced by secondary synchronized clusters of whirling oscillators. While this behavior has been observed previously, its governing conditions have not been quantitatively determined in terms of analytical criteria. Here, we develop a self-consistent theoretical framework that explicitly characterizes the secondary synchronized clusters. This analysis identifies an onset crossover mass $\tilde{m}^* \simeq 3.865$ for the emergence of secondary clusters and yields quantitative criteria for predicting both the crossover mass and the termination coupling strength at which they vanish. As a result, we determine the oscillatory regimes of coupling strengths over which global synchrony shows temporal oscillations, providing practical guidance for controlling and avoiding undesirable oscillatory behavior in inertial synchronization systems, such as power grids.

\end{abstract}


\begin{highlights}
\item Secondary clusters induce staircase-like growth of order parameter and global synchrony oscillations.
\item A self-consistent theoretical framework for secondary clusters is established.
\item An onset crossover mass $\tilde{m}^{*}\simeq3.865$ for the oscillation of secondary clusters is found.
\item Oscillatory regimes of global synchrony are quantitatively predicted.
\end{highlights}

\maketitle

\section{Introduction}\label{sec:introduction}
Synchronization is a ubiquitous phenomenon observed in a wide range of natural and artificial systems, including coordinated chemical reactions~\cite{zaikin1970concentration}, heartbeats~\cite{plonsey1987mathematical_heartmodeling}, neural activity~\cite{breakspear2010generative}, and metronome synchronization~\cite{pantaleone2002synchronization,goldsztein2021synchronization}. To understand these macroscopic collective behaviors, the Kuramoto model provides a simple yet powerful framework~\cite{kuramoto1975self,strogatz2000kuramoto,acebron2005kuramoto}. While the classical first-order Kuramoto model successfully captures the continuous phase transition in purely overdamped systems~\cite{hong2007entrainment,hong2015finite}, it is insufficient for many real-world oscillator networks that require effective inertia. In such systems, the phase evolves together with a conjugate velocity-like variable, which naturally leads to second-order Kuramoto-type equations. Starting from the classic example of synchronized Southeast Asian fireflies~\cite{ermentrout1991adaptive}, real-world realizations include power-grid networks~\cite{filatrella2008analysis,kim2018multistability}, nanoelectromechanical and optomechanical oscillator arrays with explicit mass--acceleration dynamics~\cite{matheny2019exotic,zhang2012synchronization}, and Josephson junction circuits that exhibit collective phase locking~\cite{levi1978dynamics,trees2005synchronization}. Further examples include semiconductor laser arrays and optically injected lasers, whose Lang--Kobayashi-type rate equations reduce to delayed, inertia-like phase models~\cite{kozyreff2000global,wieczorek1999unifying}, as well as physiological oscillators such as cardiac pacemaker cells, where fast voltage and slow recovery variables generate inertia-like phase behavior~\cite{glass2001synchronization,jalife1984mutual}.

The presence of inertia fundamentally alters the system's macroscopic behavior. Unlike the continuous synchronization transition seen in the first-order Kuramoto model, inertial dynamics typically exhibit a discontinuous transition accompanied by strong hysteresis between the forward and backward processes (corresponding to increasing and decreasing coupling strength, respectively)~\cite{tanaka1997first}. Theoretically analyzing these second-order dynamics is highly non-trivial because the phase and velocity variables are strongly coupled. To overcome this complexity, various analytical frameworks have been developed. The self-consistent method has been most popularly employed to understand the macroscopic order parameter by classifying the oscillator population into phase-locked and drifting groups~\cite{tanaka1997self,tanaka1997first}. Building upon this foundational approach, extensive theoretical efforts have been made to derive the exact boundaries of the hysteretic transitions~\cite{olmi2014hysteretic}, investigate characteristic behaviors under various natural-frequency distributions~\cite{olmi2016dynamics,gao2018self}, and incorporate the influence of higher-order interactions~\cite{sabhahit2024prolonged}. Despite these profound advances, the intricate non-linear dynamics of the inertial Kuramoto model, even in structurally simple all-to-all networks, leave many intermediate states and emergent patterns not fully understood.

Among these phenomena, a striking behavior has been reported in the forward process of the inertial Kuramoto model. In this process, the global synchrony exhibits a distinctive staircase-like increase as the coupling strength $K$ is raised~\cite{olmi2014hysteretic,kim2025cluster}. This non-smooth behavior, which appears through intermittent plateaus, is closely related to oscillatory dynamics of the global synchrony~\cite{olmi2014hysteretic,Belykh2016}. At a more fundamental level, this oscillatory behavior arises from the presence of whirling oscillators: in addition to the main synchronized cluster with zero phase velocity in the rotating frame, whirling oscillators form secondary synchronized clusters that rotate with nonzero but quite constant phase velocities, in contrast to fully drifting oscillators~\cite{tanaka1997first,gao2021synchronized}. This staircase behavior does not appear when the inertia $m$ is small, and even for sufficiently large $m$ it eventually disappears as $K$ increases. Despite extensive studies on hysteresis and cluster formation, the onset mass $\tilde{m}^*$ at which secondary clusters first emerge and the coupling strength $K_t$ at which they vanish for a given $m$ have not been quantitatively determined.

In this study, we propose a self-consistent analysis for the secondary synchronized clusters that governs the oscillatory behavior of the global synchrony. By constructing a theoretical framework based on an existing method~\cite{gao2018self} and by identifying the condition under which the secondary synchrony vanishes, we reveal the onset crossover mass $\tilde{m}^* \simeq 3.865$ at which secondary clusters first appear and persist for $m \gtrsim \tilde{m}^*$. For a given $K$, this condition allows us to predict the crossover mass $m^*$ at which the secondary clusters emerge. Conversely, for a given $m$, the same condition yields the termination coupling strength $K_t$, thereby enabling us to identify the oscillatory regimes of the coupling strength $K$ in which the oscillatory behavior, that is, the presence of secondary clusters, occurs in the forward process.

This paper is organized as follows. In Sec.~\ref{sec:Model_and_time_evolution}, we describe numerical observations of the oscillatory behavior of the global synchrony. Section~\ref{sec:Self-consistent_equation_of_Secondary_cluster} is devoted to establishing the mean-field equation for the secondary synchronization, solving the resulting self-consistent equation, and identifying the crossover mass and the oscillatory regime. The analysis of the forward process is presented in Sec.~\ref{sec:forward}. Finally, we discuss the results and limitations of the present study in Sec.~\ref{sec:conclusions}.


\section{Dynamics of the Inertial Kuramoto Model and Formation of Secondary Clusters}
\label{sec:Model_and_time_evolution}

\begin{figure*}[!ht]
    \centering
    \includegraphics[width=0.5\linewidth]{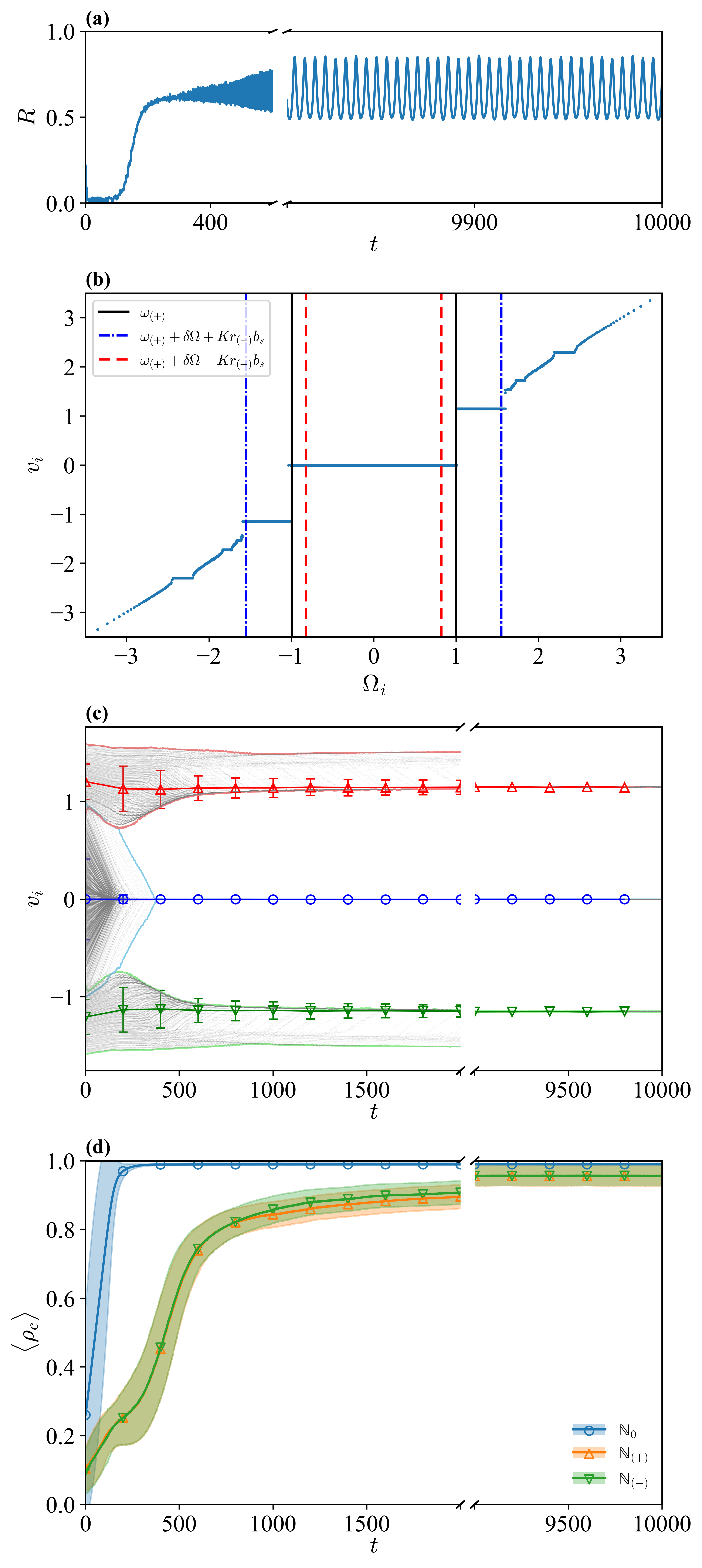}
    \caption{Formation of secondary clusters at $K=6$ and $m=6 > \tilde{m}^*$ with the marginal crossover mass $\tilde{m}^*\simeq 3.865$ (see Fig.~\ref{fig:critical_m}) for $N=5\,000$. (a) Evolution of the order parameter $R$. (b) Natural frequency $\Omega_i$ versus the temporal-averaged phase velocity $v$ for all oscillators, computed over a time window $\Delta t = 200$ at $t = 10\,000$. The black solid and blue dotted vertical lines indicate the cluster boundaries $\Omega_{(\pm),0}$ and $\Omega_{(\pm)}$, corresponding to Eq.~(\ref{eq:omega_boundaries}). The red dashed vertical line denotes the theoretical boundary separating the main cluster $\mathbb{N}_0$ from the secondary clusters $\mathbb{N}_{(\pm)}$ (see Sec.~\ref{subsec:solution}). (c) Temporal-averaged phase velocity $v$ for the oscillators belonging to each cluster. For a representative time point, the mean and standard deviation of $v$ within each cluster are shown, and the minimum and maximum values in each cluster are highlighted by colored solid lines. (d) Cluster synchrony $\rho_C$ for $\mathbb{N}_0$ and $\mathbb{N}_{(\pm)}$. The circle, triangle, and inverted triangle denote $\mathbb{N}_0$, $\mathbb{N}_\p$, and $\mathbb{N}_\m$, respectively.}
    \label{fig:r_v_evolve}
\end{figure*}
We consider the Kuramoto model with inertia for a network of $N$ coupled oscillators.
The governing equations for the system dynamics are given by a set of coupled differential equations:
\begin{equation}
m\ddot{\theta}_i + \dot{\theta}_i = \Omega_i + \frac{K}{N}\sum_{j=1}^{N} A_{ij} \sin(\theta_j - \theta_i), \quad (i = 1, \dots, N),
\label{eq:eq_of_motion}
\end{equation}
where $\theta_i$ and $m$ denote the phase and mass of oscillator $i$ ($m_i = m$ for all $N$ oscillators), and $K$ is the coupling strength. The underlying network structure is explicitly defined by the adjacency matrix elements $A_{ij}$. In this study, we focus on a globally coupled network or complete graph, where $A_{ij} = 1$ for all $i \neq j$ and $A_{ii} = 0$.
The intrinsic frequency $\Omega_i$ is regularly sampled from a Gaussian distribution $g(\Omega; \mu, \sigma)$ with zero mean ($\mu=0$) and variance $\sigma^2$, while the initial phase $\theta_i$ and phase velocity $\dot{\theta}_i$ are randomly chosen from uniform distributions over $[0, 2\pi)$.
To quantify the global synchrony at a given coupling strength $K$, we use the complex order parameter $Z$~\cite{kuramoto1984chemical}, defined as
\begin{equation}
Z(t) = R(t)e^{i\phi(t)} \equiv \frac{1}{N}\sum_j e^{i\theta_j(t)},
\label{eq:r}
\end{equation}
where $R$ measures the degree of synchrony and $\phi$ is the mean phase.
The conventional order parameter $R$ ranges from 0 to 1, with $R=0$ corresponding to complete incoherence and $R=1$ to perfect synchronization ($\theta_i=\phi$ for all $i$).

For strong coupling (large $K$), the conventional order parameter $R$ eventually reaches a large value but exhibits strong temporal fluctuations, as shown in Fig.~\ref{fig:r_v_evolve}(a).
Previous studies~\cite{tanaka1997first,gao2021synchronized} have revealed that the emergence of secondary synchronized clusters is responsible for these fluctuations.
To characterize such clusters, Ref.~\cite{gao2021synchronized} proposed defining synchronous clusters based on their mean phase velocity.
The temporal average of the phase velocity $v_i(t)$ in the steady state, over a time window $\Delta t$, is given by
\begin{equation}
v_i(t) \equiv \langle\dot{\theta_i}\rangle_t = \frac{1}{\Delta t}\int_{t}^{t+\Delta t}\dot{\theta}_i(\tau)~ {\mathrm{d}}\tau,
\label{eq:avg_v}
\end{equation}
where $\langle \cdots \rangle_t$ denotes the temporal average between $t$ and $t+\Delta t$.
Figure~\ref{fig:r_v_evolve}(b) displays $v_i(t)$ from Eq.~(\ref{eq:avg_v}) at $t=10^4$ with $\Delta t=10^2$ for the case in Fig.~\ref{fig:r_v_evolve}(a), sorted by the intrinsic frequency $\Omega_i$. Three distinct plateaus can be clearly observed, corresponding to oscillator groups (clusters) characterized by three different mean phase velocities. Oscillators on the longest plateau with $v_i=0$ in the rotating frame form the main synchronous cluster, which primarily determines the magnitude of $R$. The two shorter plateaus correspond to secondary clusters rotating at quite constant velocities in the counter-clockwise and clockwise directions, respectively, around the main cluster. The periodic motion of these secondary clusters mainly contributes to the oscillatory pattern of $R$. We denote by $\mathbb{N}_c$ the set of oscillators belonging to the synchronous cluster $c$ with mean phase velocity $\bar{v}_c$, where $c\in\{0, \p, \m\}$, $0$ represents the main cluster (thus $\bar{v}_0=0$), and $\p$ / $\m$ indicates the counter-clockwise/clockwise cluster, respectively.

The cluster formation strongly depends on the intrinsic frequency $\Omega_i$ [Fig.~\ref{fig:r_v_evolve}(b)]. The oscillators with the intrinsic frequencies around the mean value of the distribution $g(\Omega)$ primarily form the main cluster $\mathbb{N}_0$, and then the oscillator groups with the next larger magnitude of the $\Omega$ form the secondary clusters $\mathbb{N}_\both$, and the others are drifting. We denote the boundaries of the intrinsic frequencies: $\Omega_{\m,0}< \Omega_i < \Omega_{\p,0}$ for $i\in \mathbb{N}_0$, $\Omega_{\p,0}< \Omega_i < \Omega_{\p}$ for $i\in \mathbb{N}_\p$, and $\Omega_i >\Omega_\p$ for drifting oscillators. For the clockwise secondary cluster $\mathbb{N}_\m$, the boundaries $\Omega_{\m}$ and $\Omega_{\m,0}$ are also defined in a similar way to the cluster $\mathbb{N}_\p$. In other words, the sets of the synchronized oscillators are expressed as $\mathbb{N}_0=\mathbb{N}_{\Omega_{\m,0} < \Omega<\Omega_{\p,0}}, \mathbb{N}_{\p}=\mathbb{N}_{\Omega_{\p,0}<\Omega<\Omega_{\p}}$, and $\mathbb{N}_{\m}=\mathbb{N}_{\Omega_{\m}<\Omega<\Omega_{\m,0}}$, and the drifting oscillators belong to $\mathbb{N}_{\Omega>\Omega_\p} \cup \mathbb{N}_{\Omega<\Omega_\m}$. We also numerically observe that the boundaries are symmetric as $\Omega_{\p,0}\simeq -\Omega_{\m,0}$ and $\Omega_{\p}\simeq -\Omega_{\m}$ under our setting.

The trajectories of $v_i(t)$ for oscillators in $\mathbb{N}_0$, $\mathbb{N}_\p$, and $\mathbb{N}_\m$ are shown in Fig.~\ref{fig:r_v_evolve}(c).
Contrary to the rapid saturation of oscillators in the main cluster $\mathbb{N}_0$ toward $\bar{v}_0$, those in the secondary clusters $\mathbb{N}_\p$ and $\mathbb{N}_\m$ exhibit much slower relaxations toward $\bar{v}_\p$ and $\bar{v}_\m$, accompanied by relatively larger fluctuations.
These observations indicate that the clusters form in qualitatively different ways over time. The cluster synchrony $\rho_c$ quantifies these temporal patterns:
\begin{equation}
\rho_{c}(t)e^{i\phi_c(t)}=\frac{1}{N_c}\sum_{j\in \mathbb{N}_c} e^{i\theta_j(t)},
\label{fig:r_cluster}
\end{equation}
where $N_c = |\mathbb{N}_c|$ denotes the number of oscillators in the set $\mathbb{N}_c$.
We observe that the main cluster rapidly reaches almost complete synchrony ($\rho_0 \approx 1$), while the onsets of the secondary synchronizations and reaching $\rho_{(\pm)}\approx 1$ are delayed until the relaxation of the main cluster is nearly finished [Fig.~\ref{fig:r_v_evolve}(d)]. In other words, the coexistence of the secondary clusters rotating at opposite velocities contributes to the oscillating pattern of $R$.

\begin{figure}[h]
    \centering
    \includegraphics[clip,width=0.7\linewidth]{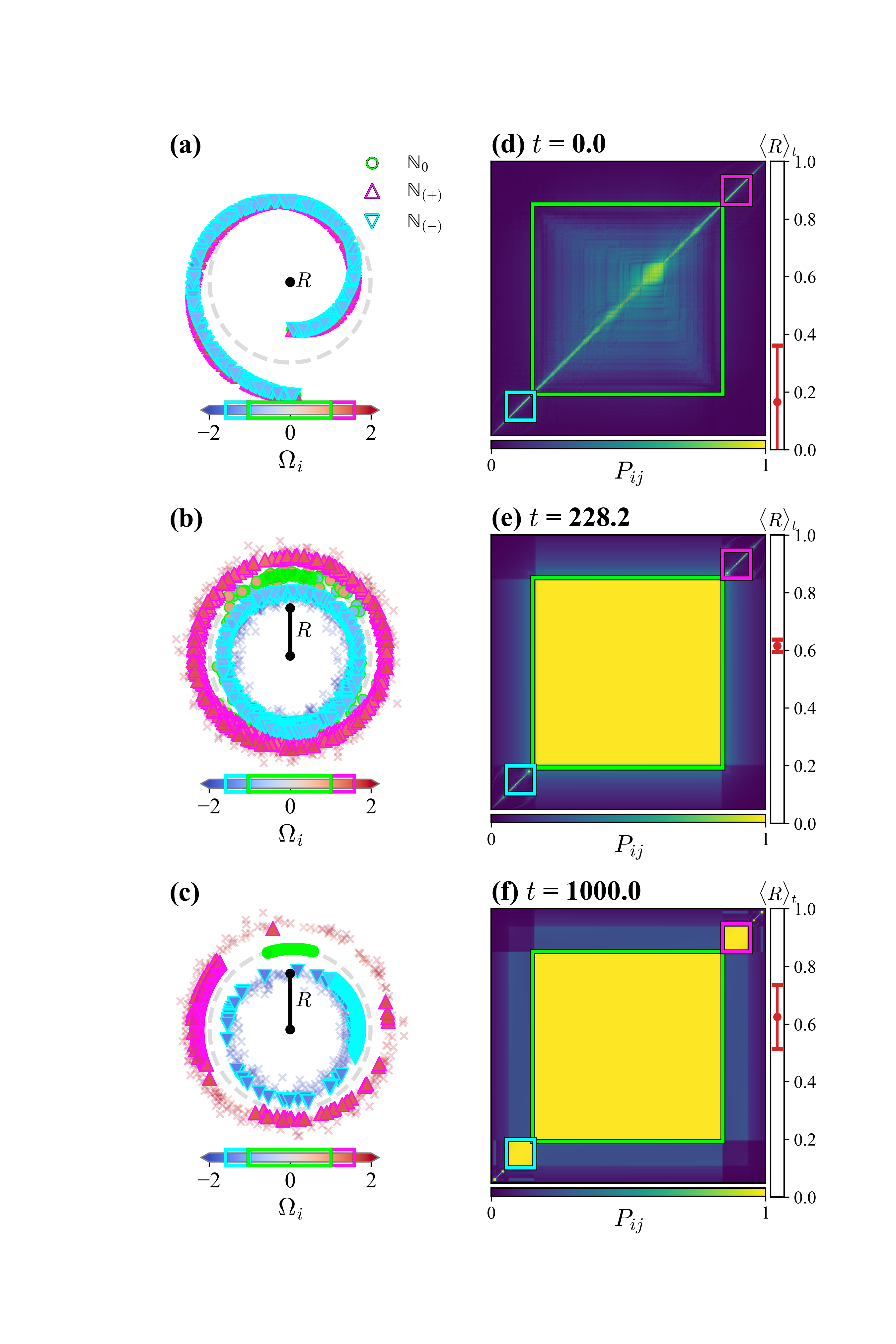}
    \caption{Interaction of oscillators for $m=6$, $K=6$, and $N=5\,000$ with a temporal averaging window $\Delta t = 200$ at three representative times: (a, b) the initial state, (c, d) the stage where $\rho_0 \approx 1$ but $\rho_{(\pm)} < 1$, and (e, f) the stage where both $\rho_0 \approx 1$ and $\rho_{(\pm)} \approx 1$. The left panels (a–c) show $\theta_i$ and $\dot{\theta}_i$ in a rotating frame, where each point represents an oscillator with angular coordinate $\theta_i$ and radial coordinate $\dot{\theta}_i$. The gray dashed circle indicates $\dot{\theta}_i = 0$; larger (smaller) radial distance corresponds to more positive (negative) phase velocity. Colors denote intrinsic frequencies $\Omega_i$. Symbols and colored borders mark the pre-defined clusters from Fig.~\ref{fig:r_v_evolve}(b): green circles for $\mathbb{N}_0$, pink triangles for $\mathbb{N}_\p$, and cyan inverted triangles for $\mathbb{N}_\m$. The magnitude of the global synchrony $R$ is also shown with mean phase $\phi = 0$. The right panels (d–f) display the interaction matrix $P_{ij}$, sorted by intrinsic frequency, with colored boxes indicating the cluster boundaries.
}
    \label{fig:Pij}
\end{figure}


To clarify the oscillators' temporal behaviors and cluster formation, we arrange the $N$ oscillators by their phase $\theta_i$ and velocity $\dot{\theta}_i$, as shown in Figs.~\ref{fig:Pij}(a),~\ref{fig:Pij}(b),~and~\ref{fig:Pij}(c). An oscillator $i$'s position is represented as $(\dot{\theta}_i, \theta_i)$ in polar coordinates (i.e., radius $\dot{\theta}_i$ and angle $\theta_i$). The radius of the dashed circle denotes the reference velocity of the rotating frame, so a larger/smaller radius than the reference velocity indicates motion in the counter-clockwise/clockwise direction, respectively. We pre-determine the clusters $\mathbb{N}_0, \mathbb{N}_\p$, and $\mathbb{N}_\m$ to which the oscillators belong in the steady state, as identified in Fig.~\ref{fig:r_v_evolve}(b).

At $t=0$, all oscillators start to move from zero velocity $v_i=0$ (the adjusted reference velocity) and are uniformly distributed according to the initial condition of phases [Fig.~\ref{fig:Pij}(a)]. Figure~\ref{fig:Pij}(b) shows the oscillators at $t=228.2$, when the main cluster is almost synchronized ($\rho_0 \approx 1$) and the onsets of the secondary synchronizations appear ($\rho_{\both} \approx 0$). Most oscillators labeled as the main cluster are synchronized with $v_i=\bar{v}=0$ for $i\in\mathbb{N}_0$, except for a few oscillators. The others are still desynchronized, but some of them begin to cluster in terms of velocity depending on the sign of the intrinsic frequency $\Omega_i$: $v_i<0$ for $\Omega_i<0$ and $v_i>0$ for $\Omega_i>0$. Finally, in the steady state, the main cluster and the secondary clusters are firmly formed with $\bar{v}_0=0$, $\bar{v}_{\p} > 0$, and $\bar{v}_{\m} < 0$.

To clearly identify the cluster formation, we propose a visualization method that presents the interactions between oscillators, defining $P_{ij}$ between two oscillators $i$ and $j$ based on their phase synchrony as
\begin{equation}
    P_{ij}(t) = \frac{1}{\Delta t} \left|\int_{t}^{t+\Delta t} e^{i\left[\theta_j(\tau)-\theta_i(\tau)\right]} {\mathrm{d}}\tau\right|,
\label{eq:Pij}
\end{equation}
and note that $P_{ij} = P_{ji}$ and $P_{ii}=1$. The snapshots of the interaction matrix are shown in Fig.~\ref{fig:Pij} at representative times, along with the temporal average $\langle R\rangle_t$ of the global synchrony in Eq.~(\ref{eq:r}) and its standard deviation $\sigma_t=\sqrt{\langle ( R(t)- \langle R \rangle_t )^2 \rangle_t }$. The oscillator indices are sorted by intrinsic frequency, with $\Omega_i$ on the horizontal axis and $\Omega_j$ on the vertical axis, consistent with Fig.~\ref{fig:r_v_evolve}(b).

The initial $P_{ij}$ remains low with low $\langle R\rangle_t$ and high $\sigma_t$, as expected [Fig.~\ref{fig:Pij}(d)]. In Fig.~\ref{fig:Pij}(e), at $t=228.2$, the oscillators in $\mathbb{N}_0$ interact maximally, i.e., $P_{ij}\approx 1$. Meanwhile, strong interactions in $\mathbb{N}_\p$ or $\mathbb{N}_\m$ emerge around the oscillators whose $\Omega_i$ values are close to those in the main cluster. The oscillators in the secondary clusters are still not clearly distinguished from the drifting oscillators, leading to partial global synchrony $\langle R\rangle_t\approx 0.62$ with small $\sigma_t$. In the steady state, a comparable number of oscillators in the secondary clusters interact strongly [Fig.~\ref{fig:Pij}(f)]. The global synchrony is similar to that in the transient period shown in Fig.~\ref{fig:Pij}(e) but exhibits larger fluctuations. From these results, one can infer that the oscillating and fluctuating behavior of the global synchrony $\langle R\rangle_t$ in the steady state is closely related to the formation of secondary clusters of sizes comparable to the system size.

\section{Self-Consistent Theory of Secondary Synchronization}
\label{sec:Self-consistent_equation_of_Secondary_cluster}
To understand the effect of the secondary clusters theoretically, we establish and solve the self-consistent equation for the secondary synchrony by extending the framework suggested in previous studies~\cite{tanaka1997first,olmi2014hysteretic,gao2018self, gao2021synchronized}. This analysis deepens our understanding of the system by identifying the crossover mass at which the secondary cluster emerges and by determining the persistence of the oscillating state of the global synchrony.

\subsection{Establishing the Mean-Field Framework}
\label{subsec:establishemnt}

\begin{figure*}
    \centering
    \includegraphics[width=1\linewidth]{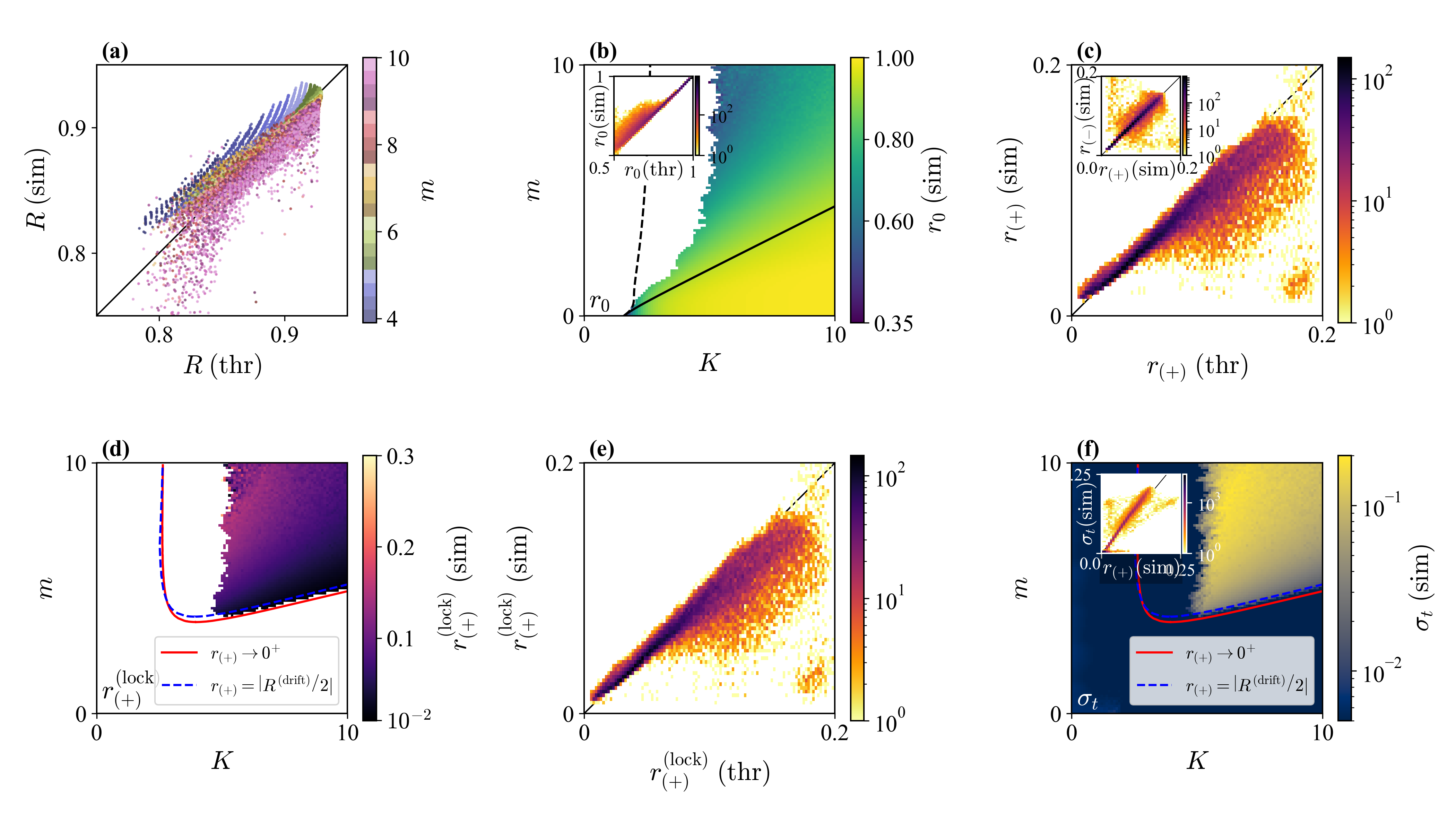}
    \caption{
    Comparison between the theoretical analysis and numerical simulation. We use $N=5\,000$ oscillators on the $K\text{--}m$ plane ($0 \le K, m \le 10$) with $\mathrm{d}K=\mathrm{d}m=0.1$, evaluated at $t = 10\,100$ with a temporal-averaging window $\Delta t = 50$. (a) Scatter plot comparing the global order parameter $R$ obtained from simulation and theory [Eq.~(\ref{eq:z_decompose})] for the parameter regime where secondary clusters exist. The color indicates the mass, and the solid line denotes $y=x$. (b) Phase diagram of the main-cluster synchrony $r_0$ for $\mathbb{N}_0$ on the parameter space. The dashed and solid curves mark the theoretical critical boundaries obtained from the self-consistent equation [Eq.~(\ref{eq:z_lock_drift})]: the minimum $K_{\rm min}$ required to sustain nonzero $R$ (dashed) and the minimum $K_c$ at which $R$ first becomes nonzero (solid) [Fig.~\ref{fig:forward}(b)]. Inset: density plot comparing simulation and theory, with $y=x$ (solid line). (c) Density plot of the secondary synchrony $r_\p$ from simulation and theory [Eq.~(\ref{eq:rplus_selfcon})], with $y=x$ marked by the solid line. Inset: simulation density plot of $r_\p$ and $r_\m$. (d) Phase diagram of the secondary-cluster synchrony $r^{\lc}_\p$ for $\mathbb{N}_\p$ on the parameter space. The dashed and solid curves represent the predicted onset boundaries of secondary synchronization given by $r_\p = |R^{\dr}/2|$ and $r_\p \to 0^{+}$, respectively. The white regime indicates the absence of the secondary cluster in simulations. The white region inside the boundary corresponds to a theoretically allowed but numerically infeasible regime due to the prerequisite condition $R>0$.} 
(e) Density plot comparing $r^{\lc}_\p$ between simulation and the theoretical expression [Eq.~(\ref{eq:rplus_lock})], with the solid line showing $y=x$.
(f) Temporal fluctuation $\sigma_t$ of the global synchrony $R$ from simulation. Inset: density plot of $\sigma_t$ versus $r_\p$, with the theoretical line $y = \sqrt{2 - \left({\pi}/{8}\right)^2 - \left({\pi}/{4}\right)^4}\, x$ from Eq.~(\ref{eq:sigma_rp}).
    \label{fig:sim_thr}
\end{figure*}
It is well known that the Kuramoto model with inertia in Eq.~(\ref{eq:eq_of_motion}) is reduced to the mean-field equation by using the complex order parameter [Eq.~(\ref{eq:r})]:
\begin{equation}
    m\ddot{\theta}_i +\dot{\theta}_i=\Omega_i +KR\sin(\phi -\theta_i).
\label{eq:mf_eq}
\end{equation}
The quantities $R$, $\phi$, and $\theta_i$ are explicitly time-dependent [e.g., $R(t)$], but we omit the notation $t$ for convenience in Eq.~(\ref{eq:mf_eq}). To derive the self-consistent equation for the secondary clusters, we decompose the complex order parameter $Z$~\cite{gao2021synchronized} as
\begin{align}
    Z &= \frac{1}{N}\left( \sum_{j\in\mathbb{N}_0} + \sum_{j\in\mathbb{N}_{\Omega>\Omega_{\p,0}}} + \sum_{j\in\mathbb{N}_{\Omega<\Omega_{\m,0}}}\right)e^{i\theta_j} \nonumber\\
      &= z_0 + z_\p + z_\m ~,
\label{eq:z_decompose}
\end{align}
where the complex order parameter $z_c$ is defined as $z_c \equiv r_c e^{i\phi_c} \equiv \frac{1}{N} \sum_{j\in \mathbb{N}_c^{\prime}} e^{i \theta_j}$, with the oscillator sets $\mathbb{N}_c^{\prime}=\mathbb{N}_0$ for $z_0$ and $\mathbb{N}_{\Omega\gtrless\Omega_{(\pm),0}}$ for $z_{\both}$. The last two terms on the right-hand side of Eq.~(\ref{eq:z_decompose}) correspond to the union of the oscillators in the secondary clusters $\mathbb{N}_\both$ and the drifting oscillators with $\Omega \gtrless \Omega_{(\pm),0}$. Equation~(\ref{eq:z_decompose}) can then be written as
\begin{equation}
    R e^{i\phi} = r_0 e^{i\phi_0} + r_\p e^{i\phi_\p} + r_\m e^{i\phi_\m}.
    \label{eq:Rphi_decompose}
\end{equation}
The mean phase of group $c$ takes the form $\phi_{c} = \omega_c t + \Phi_c$, where $\omega_c$ and $\Phi_c$ denote the rotating frequency and phase of that group, respectively. To formalize these observations, we now construct a theoretical description based on the mean-field equation.

For the sake of solvability, we focus on the symmetric case defined as follows. The global mean phase $\phi$ is determined solely by the mean phase $\phi_0$ of the main cluster, i.e., $\phi = \phi_0 = 0$ in the rotating frame in which the global mean phase, dominated by the main synchronized cluster, is stationary. We assume identical synchrony and opposite rotating frequencies for the secondary clusters, i.e., $r_\p = r_\m$ and $\omega_\p=-\omega_\m$, which can be realized by the regular sampling of $\Omega_i$ in the numerical simulations of Sec.~\ref{sec:Model_and_time_evolution}. The assumption $r_\p=r_\m$ used in the theory is well reproduced in the numerical results, as seen in the inset of Fig.~\ref{fig:sim_thr}(c). Under this restriction, the real and imaginary parts of the complex order parameter $Z$ are written as
\begin{align}
    \Re(Z) = R &= r_0 \cos \phi_0 + r_\p \cos \phi_\p + r_\m \cos \phi_\m \nonumber \\
    &= r_0 + r_\p \cos\left(\omega_\p t+\Phi_\p \right)+ r_\p \cos \left(-\omega_\p t+\Phi_\m \right), \label{eq:re_z}
\end{align}
and
\begin{align}
    \Im(Z)= 0 &= r_\p \sin \phi_\p + r_\m \sin \phi_\m \nonumber\\
        &= r_\p \sin\left(\omega_\p t+\Phi_\p \right) + r_\p \sin \left(-\omega_\p t+\Phi_\m \right). \label{eq:im_z}
\end{align}
We start from $R(t=0)=r_0(t=0)$, which yields $\Phi_\p = -\Phi_\m = -\pi/2$ from Eqs.~(\ref{eq:re_z}) and~(\ref{eq:im_z}). The mean phases of the secondary clusters are thus rearranged as $\phi_\p=\omega_\p t-{\pi / 2}$ and $\phi_\m=-\omega_\p t+{\pi / 2}=-\phi_\p$. This symmetric condition implies a phase difference of $\pi$ between the secondary clusters, i.e., $|\Phi_\p-\Phi_\m| = \pi$. To formulate the self-consistent equation for the secondary cluster (specifically for $r_\p$), we apply a coordinate transformation $\theta_i \to \theta_i+\phi_\p$. Substituting these expressions into the mean-field equation in Eq.~(\ref{eq:mf_eq}), we obtain
\begin{align} 
    m\ddot{\theta}_i +\dot{\theta}_i &=\Omega_i- \omega_\p +K r_0\cos{\left(\omega_\p t+\theta_i\right)} - Kr_\p \sin \theta_i  + K r_\p \sin \left(2\omega_\p t+\theta_i\right).
    \label{eq:mf_eq_rplus}
\end{align}
The two terms $\cos\left(\omega_\p t+\theta_i\right)$ and $\sin\left(2\omega_\p t+\theta_i\right)$ on the right-hand side represent the relative rotations of the main cluster $\mathbb{N}_0$ and the set of clockwise oscillators $\mathbb{N}_{\Omega<\Omega_{\m,0}}$ (including $\mathbb{N}_\m$) with respect to the counter-clockwise secondary cluster $\mathbb{N}_\p$.

The explicitly time-dependent terms on the right-hand side of Eq.~(\ref{eq:mf_eq_rplus}) can be averaged out using $\langle e^{i \theta}\rangle_{t; \Delta t=T}={ \frac{1}{T} }\int_{t}^{t+T} e^{i \theta(\tau)} {\mathrm{d}}\tau$~\cite{gao2018self}, where the time window $\Delta t$ is chosen as the period $T$ of the limit cycle. In the spirit of Ref.~\cite{gao2018self}, Eq.~(\ref{eq:mf_eq_rplus}) is converted into
\begin{align}
    m\ddot{\theta}_i +\dot{\theta}_i & \approx \Omega_i - \omega_\p + \left\langle K r_0\cos{\left(\omega_\p t+\theta_i\right)} \right\rangle - K r_\p \sin \theta_i + \left\langle K r_\p \sin \left(2\omega_\p t+\theta_i\right) \right\rangle \nonumber\\
    & = \Omega_i - \omega_\p - \delta \Omega - K r_\p \sin \theta_i~,
    \label{eq:mf_eq_rplus_average}
\end{align}
where $-\delta\Omega\equiv \left\langle K r_0\cos{\left(\omega_\p t+\theta_i\right)} \right\rangle + \left\langle K r_\p \sin \left(2\omega_\p t+\theta_i\right) \right\rangle$. Note that Eq.~(\ref{eq:mf_eq_rplus_average}) is formally dependent on $i$ through $\theta_i$, which enters only as a phase shift since the period is governed by $\omega_{(+)}$ and can thus be treated as independent of $i$. Equation~(\ref{eq:mf_eq_rplus_average}) can be rewritten in the simple form of a second-order differential equation with a constant torque, $\ddot{\theta}+a\dot{\theta}=b-\sin\theta$, using a rescaled time $\tau=\sqrt{\frac{K r_\p}{m}}t$, which has a limit-cycle solution. The rescaled parameters $a$ and $b$ are defined as:
\begin{subequations}
    \begin{equation}
    a = \frac{1}{\sqrt{K r_\p m}},
    \label{eq:a}
\end{equation}
\begin{equation}
    b = \frac{\Omega-\omega_\p-\delta\Omega}{K r_\p}.
    \label{eq:b}
\end{equation}
\label{eq:a_b}
\end{subequations}


\subsection{Brief Review of Temporal Averaging and the Self-Consistent Equation for Global Synchrony}
\label{subsec:solution}

Our main goal is to derive and solve a closed self-consistent equation for $r_\p$.
In doing so, we briefly review a general method for obtaining the solution to the differential equation $\ddot{\theta}+a\dot{\theta}=b-\sin\theta$ proposed by Ref.~\cite{gao2018self} in this Section, to provide a self-contained analysis.
The main concept of this method is to evaluate the global order parameter (corresponding to $Z$ or $R$ in this study) by distinguishing oscillators into two distinct dynamical states: a locked group (forming a stable fixed point) and a drifting group (moving along a limit cycle). By computing the temporal average of the fast-oscillating phases for the drifting group over a full period, the explicitly time-dependent dynamics are systematically mapped to a mathematically tractable form.

The stable solution is a fixed point for $0<b<b_S(a)$ and a limit-cycle solution for $b>b_L\equiv 1$.
In the intermediate regime $b_S(a) < b < b_L$, the system exhibits bistability characterized by the coexistence of a stable fixed point and a limit cycle.
The higher-order approximation $b_S(a)\approx 4\pi^{-1} a - 0.3056 a^3$ for $a\lesssim 1.193$ was obtained in Ref.~\cite{Belykh2016}, and $b_L\geq b_S$ (see the phase diagram in Fig.~1 of Ref.~\cite{gao2018self}).
For the limit cycle caused by the drifting group, the average $\langle e^{i\theta}\rangle$ can be approximated to leading order as $\langle e^{i\theta}\rangle \approx \frac{1}{2}\left(-1+i\frac{a^2}{b}\right)\frac{a^2}{a^4+b^2}$.
Using this expression of temporal average for the drifting group, the authors finally constructed the self-consistent equation for the global order parameter $Z=Z^{\lc}+Z^{\dr}$ under the forward and backward processes as
\begin{subequations}
    \begin{equation}
        Z^{\lc}=\int_{|b|<b_P(a)} G(b) \left[\sqrt{1-b^2} + ib\right] \,{\mathrm{d}}b ~, \label{eq:z_lock}
    \end{equation}
    \begin{equation}
        Z^{\dr}=\int_{|b|>b_P(a)} G(b) \left[ \frac{1}{2} \left(-1+i\frac{a^2}{b}\right) \frac{a^2}{a^4+b^2} \right] \,{\mathrm{d}}b ~, \label{eq:z_drift}
    \end{equation}
    \label{eq:z_lock_drift}
\end{subequations}
where $G(b)$ is the transformed form of the density $g(\Omega)$ under the change of variables from $\Omega$ to $b$, i.e., $G(b)\,{\mathrm{d}}b = g(\Omega)\,{\mathrm{d}}\Omega$. 
The theoretical framework of Ref.~\cite{gao2018self} is valid for general forms of $g(\Omega)$, including asymmetric distributions. In our analysis, we will use the numerically obtained $g(\Omega)$ resulting from the regular sampling. Here $b_P=b_S$ for the forward process or $b_P=b_L$ for the backward process.
Separating the real and imaginary parts of Eq.~(\ref{eq:z_lock_drift}) yields the self-consistent equation for the global synchrony $R$, i.e., $R^{\lc}=\Re(Z^{\lc})$ (corresponding to the main-cluster synchrony $r_0$) and $R^{\dr}=\Re(Z^{\dr})$.
To confirm the validity of this framework, we plot the theoretical result for $R$, which agrees with the simulation result as shown in Fig.~\ref{fig:sim_thr}(a).
The phase diagram for the main-cluster synchrony $r_0$ is displayed in Fig.~\ref{fig:sim_thr}(b).
The discrepancy of the phase boundary between the simulation and theory [$K_{\rm min}$ and $K_c$;
see the caption in Fig.~\ref{fig:sim_thr}(b) and Fig.~\ref{fig:forward}(a)] is caused by the finite-size effect, and the similar deviations have already been reported in previous studies~\cite{olmi2014hysteretic, gao2018self}.

\subsection{Extension of the Self-Consistent Equation to Secondary Clusters}
\label{subsec:secondary_cluster_eq}
We modify the formulation for the global synchrony $R$ in Sec.~\ref{subsec:solution} to focus on the synchrony $r_\p$ of the secondary cluster. Let us declare in advance that we will map $Z^{\rm(lock)}\to z_{(+)}^{\rm(lock)}$ and $Z^{\rm(drift)}\to z_{(+)}^{\rm(drift)}$ in the rotating frame from the viewpoint of the secondary cluster. After constructing the self-consistent equation for $z_{(+)}$ similar to Eq.~(\ref{eq:z_lock_drift}), we finally obtain the secondary synchrony $r_{(+)}=\Re\left(z_{(+)}\right)$.

We apply the self-consistent method in Eq.~(\ref{eq:z_lock_drift}) to the mean-field equation for the secondary cluster in Eq.~(\ref{eq:mf_eq_rplus_average}). From the viewpoint of the counter-clockwise secondary cluster, the drifting groups correspond to the main cluster and the clockwise secondary cluster, and their effect is represented by $\delta\Omega$. To evaluate $\delta\Omega$, we adopt the temporal averaging method reviewed in Sec.~\ref{subsec:solution}, using the coefficients $a$ and $b$ in Eq.~(\ref{eq:a_b}). Since the coefficient $b$ already contains $\delta\Omega$, the equation must in principle be solved self-consistently with respect to $\delta\Omega$. To obtain an explicit closed-form expression, we therefore employ a leading-order approximation corresponding to the zeroth-order expansion in $\delta\Omega$. We have also examined higher-order corrections in $\delta\Omega$ (although the derivation is not shown). Including higher-order terms does not alter the main finding of this work, which will be discussed later. In addition, because the temporal averaging is performed over a period dominantly governed by the cluster velocity, the intrinsic frequency $\Omega$ entering $b$ mainly modifies the phase offset of the periodic motion relative to the cluster period. With these considerations, the temporal average can be written as
\begin{equation}
    \langle e^{i \left(\omega_\p t + \theta\right)} \rangle \approx -\frac{1}{2} \left( m+\frac{i}{\omega_\p} \right)\frac{K r_\p}{1+\omega_\p^2 m^2}~.
\label{eq:theta_average}
\end{equation}
The real part on the left-hand side of Eq.~(\ref{eq:theta_average}) becomes $\langle \cos \left(\omega_\p t+\theta \right)\rangle$, and the imaginary part with $\omega_\p \to 2\omega_\p$ becomes $\langle \sin \left(2\omega_\p t+\theta \right)\rangle$. Substituting the results of Eq.~(\ref{eq:theta_average}) into Eq.~(\ref{eq:mf_eq_rplus_average}) through $\delta\Omega$, the mean-field equation for the secondary cluster, expressed in terms of $r_\p$, becomes
\begin{equation}
    m \ddot{\theta}_i + \dot{\theta}_i = \Omega_i - \omega_\p 
    - K r_\p \left[
        \frac{m K r_0}{2\left(m^2 \omega_\p^2 + 1\right)}
        + \sin\theta_i
        + \frac{K r_\p}{4\omega_\p \left(4\omega_\p^2 m^2 + 1\right)}
    \right].
    \label{eq:final_mf_rplus}
\end{equation}
The solution of Eq.~(\ref{eq:final_mf_rplus}) in terms of $r_\p$ can be obtained by mapping it onto Eq.~(\ref{eq:z_lock_drift}) after decomposing $z_\p = z^{\lc}_{\p} + z^{\dr}_{\p}$. The oscillators in $\mathbb{N}_\p$ are \emph{locked} from the viewpoint of $z_\p$, which enables us to express $z_\p^{\lc}$. The remaining oscillators belonging to $\mathbb{N}_{\Omega>\Omega_{\p}}$ are regarded as \emph{drifting} oscillators, corresponding to $z_\p^{\dr}$~\cite{olmi2014hysteretic,tanaka1997self,gao2018self}.

Compared with Eq.~(\ref{eq:z_lock_drift}), we replace $Z\to z_\p$ and use 
$G(b)=g(\Omega)\left(\frac{{\mathrm{d}}b}{{\mathrm{d}}\Omega}\right)^{-1}=K r_\p \, g(\Omega)$. 
Under this modification, the locking and drifting conditions become  
$\left| \frac{\Omega-\omega_\p - \delta\Omega}{K r_\p} \right| < b_P$ for $z_\p^{\lc}$ and  
$ \frac{\Omega-\omega_\p - \delta\Omega}{K r_\p} > b_P$ for $z_\p^{\dr}$  
(i.e., one side of the range $\left|\frac{\Omega-\omega_\p - \delta\Omega}{K r_\p} \right| > b_P$). 
Applying the change of variables ($b\to\Omega$), these conditions translate into  
$\Omega_1 < \Omega < \Omega_2$ for $z_\p^{\lc}$ and $\Omega > \Omega_2$ for $z_\p^{\dr}$, 
with  
$\Omega_1=\omega_\p + \delta\Omega - K r_\p b_P$  
and  
$\Omega_2=\omega_\p + \delta\Omega + K r_\p b_P$. 
These quantities should correspond to the boundary frequencies $\Omega_{\p,0}$ and $\Omega_\p$.

In addition to this theoretical mapping, numerical observations reveal that 
$\omega_\p > \omega_\p + \delta\Omega - K r_\p b_P$ 
in our parameter regime, and that oscillators forming the secondary cluster $\mathbb{N}_\p$ consistently appear only for $\Omega > \omega_\p$. 
Based on this observation, we redefine the lower boundary as $\Omega_1 \equiv \omega_\p$ [Fig.~\ref{fig:r_v_evolve}(b)]. 
Consequently, the theoretical boundary frequencies consistent with our numerical detection become
\begin{subequations}
    \begin{equation}
        \Omega_{\p,0} = \Omega_1 = \omega_\p~, \label{eq:omega_p0}
    \end{equation}
    \begin{equation}
        \Omega_\p = \Omega_2 = \omega_\p + \delta\Omega + K r_\p b_P~, \label{eq:omega_p}
    \end{equation}
    \label{eq:omega_boundaries}
\end{subequations}
which are fully determined once a theoretical expression for $\omega_\p$ is specified. 
We have numerically verified the consistency of these results (data omitted for brevity).

The above observation tells us that the oscillators in the main cluster $\mathbb{N}_0$ have intrinsic frequencies in the range $-\omega_\p < \Omega < \omega_\p$. In the context of Eq.~(\ref{eq:z_lock_drift}), this implies the relation 
$K R\, b_P(a') = \omega_\p$, with $a' \equiv {1}/{\sqrt{K R m}}$. Hence,
\begin{equation}
    \omega_\p = 
    \begin{cases}
        \frac{4}{\pi}\sqrt{\frac{K R}{m}} - 0.3056\sqrt{\frac{1}{K R m^3}}, & \text{for a forward process } (b_P=b_S), \\
        K R, & \text{for a backward process } (b_P=b_L=1),
    \end{cases}
    \label{eq:lowerlimit}
\end{equation}
where the expression for the backward process corresponds to the branch starting from the fully synchronized state ($R = 1$)~\cite{gao2018self}. Due to this specificity of the backward process, we confine our interest to the forward process in this study.

Since the simulations in Figs.~\ref{fig:r_v_evolve} and~\ref{fig:Pij} correspond to the forward process (due to the initial condition $R(0)=0$), we employ $b_P=b_S$ and obtain the self-consistent equations for $r_\p^{\lc}$ and $r_\p^{\dr}$:
\begin{subequations}
    \begin{equation}
        r_\p^{\lc} = \int_{\Omega_1}^{\Omega_2} g(\Omega) 
        \sqrt{1 - \left(\frac{\Omega - \omega_\p - \delta\Omega}{K r_\p}\right)^2} \, {\mathrm{d}}\Omega~, 
        \label{eq:rplus_lock}
    \end{equation}
    \begin{equation}
        r_\p^{\dr}
        = -\frac{K r_\p m}{2} 
        \int_{\Omega_2}^{\infty} 
        \frac{g(\Omega)}{1 + m^2\left(\Omega - \omega_\p - \delta\Omega\right)^2} \, {\mathrm{d}}\Omega ~.
        \label{eq:rplus_drift}
    \end{equation}
    \label{eq:rplus_lock_drift}
\end{subequations}

The drifting oscillators move predominantly with their intrinsic speeds $\Omega$, so we approximate $\Omega - \omega_\p - \delta\Omega \approx \Omega$ in Eq.~(\ref{eq:rplus_drift}). Summing the locked and drifting contributions finally gives the self-consistent equation for $r_\p$:
\begin{align}
    r_\p = r_\p^{\lc} + r_\p^{\dr}
    \approx & \int_{\omega_\p}^{\omega_\p + \delta\Omega + K r_\p b_S(a)} 
    g(\Omega)\sqrt{1-\left(\frac{\Omega - \omega_\p - \delta\Omega}{K r_\p}\right)^2} \, {\mathrm{d}}\Omega 
    \nonumber \\
    & -\frac{K r_\p m}{2} 
    \int_{\omega_\p + \delta\Omega + K r_\p b_S(a)}^{\infty} 
    g(\Omega)\frac{1}{1+m^2\Omega^2} \, {\mathrm{d}}\Omega ~.
    \label{eq:rplus_selfcon}
\end{align}
A good agreement is found between the theoretical prediction of $r_\p$ and the simulation results in Fig.~\ref{fig:sim_thr}(c). When the mass becomes large, tertiary or higher-order clusters may appear~\cite{kim2025cluster}, which are not captured by our theoretical framework, leading to a slight underestimation in the simulations. The phase diagram for the secondary clusters is displayed in Fig.~\ref{fig:sim_thr}(d). 
The synchronization regime for the secondary clusters is narrower than that of the global synchrony 
[Fig.~\ref{fig:sim_thr}(b)], meaning that the main cluster 
must synchronize prior to the secondary clusters~\cite{kim2025cluster}. The theoretical form $r_\p^{\lc}$ in 
Eq.~(\ref{eq:rplus_lock}) agrees well with the numerical results, as shown in 
Fig.~\ref{fig:sim_thr}(e). Furthermore, from Figs.~\ref{fig:sim_thr}(c) and~\ref{fig:sim_thr}(e), 
we observe that $r_\p \simeq r_\p^{\lc}$ with only a very small contribution from $r_\p^{\dr}$.


From the numerical observation in Fig.~\ref{fig:r_v_evolve}, we infer that the emergence of secondary clusters governs the oscillating behavior of the global synchrony $R$, motivating us to investigate its temporal fluctuation. 
The temporal average of $R$ is computed by averaging Eq.~(\ref{eq:re_z}):
\begin{align}
    \langle R\rangle_t &= \langle r_0 \cos \phi_0 \rangle_t 
    + \langle r_\p \cos\phi_\p \rangle_t 
    + \langle r_\m \cos\phi_\m \rangle_t \nonumber \\
    &\approx r_0 - \frac{m K r_0 r_\p}{m^2 \omega_\p^2 + 1},
\label{eq:temporal_mean_rt}
\end{align}
where the last approximation uses the evaluation of the oscillating quantity in Eq.~(\ref{eq:theta_average}).
The temporal fluctuation over one period is estimated through the standard deviation $\sigma_t$, obtained from Eqs.~(\ref{eq:z_decompose}) and~(\ref{eq:temporal_mean_rt}) as
\begin{align}
    \sigma_t^2 
    &= \left\langle \left( R - \langle R\rangle_t \right)^2 \right\rangle_t \nonumber \\
    &= 4 r_\p^2 \left( \langle \cos^2 \phi_\p \rangle_t 
       - \langle \cos \phi_\p \rangle_t^2 \right) \nonumber \\
    &\approx 
    4 r_\p^2 \left[ 
        \frac{1}{2}
        - \frac{1}{4} \frac{m K r_0}{4 m^2 \omega_\p^2 + 1}
        - \left( \frac{1}{2} \frac{m K r_0}{m^2 \omega_\p^2 + 1} \right)^2
    \right]~.
\label{eq:temporal_std_rt}
\end{align}
The qualitative behaviors of the secondary synchrony $r_\p^{\lc}$ and the global fluctuation $\sigma_t$ resemble each other in Figs.~\ref{fig:sim_thr}(d) and~\ref{fig:sim_thr}(f). 
For large $m$ and $K$, we have $a' < 1$ and therefore 
$\omega_\p \approx \frac{4}{\pi}\sqrt{\frac{K R}{m}}$ from Eq.~(\ref{eq:lowerlimit}). 
Assuming $r_0 \approx R$, the fluctuation in Eq.~(\ref{eq:temporal_std_rt}) reduces to
\begin{equation}
    \sigma_t^2 \approx 
    \left[
        2 - \left(\frac{\pi}{8}\right)^2 - \left(\frac{\pi}{4}\right)^4
    \right] r_\p^2~.
    \label{eq:sigma_rp}
\end{equation}
This relation between $r_\p$ and $\sigma_t$ clearly demonstrates that the temporal fluctuation of $R$ depends on the presence of the secondary clusters, consistent with the numerical result shown in the inset of Fig.~\ref{fig:sim_thr}(f).

\subsection{Crossover Mass $m^*$ for Secondary Synchronization}
\label{subsec:critical_mass}

\begin{figure}[h]
    \centering
    \includegraphics[width=0.5\linewidth]{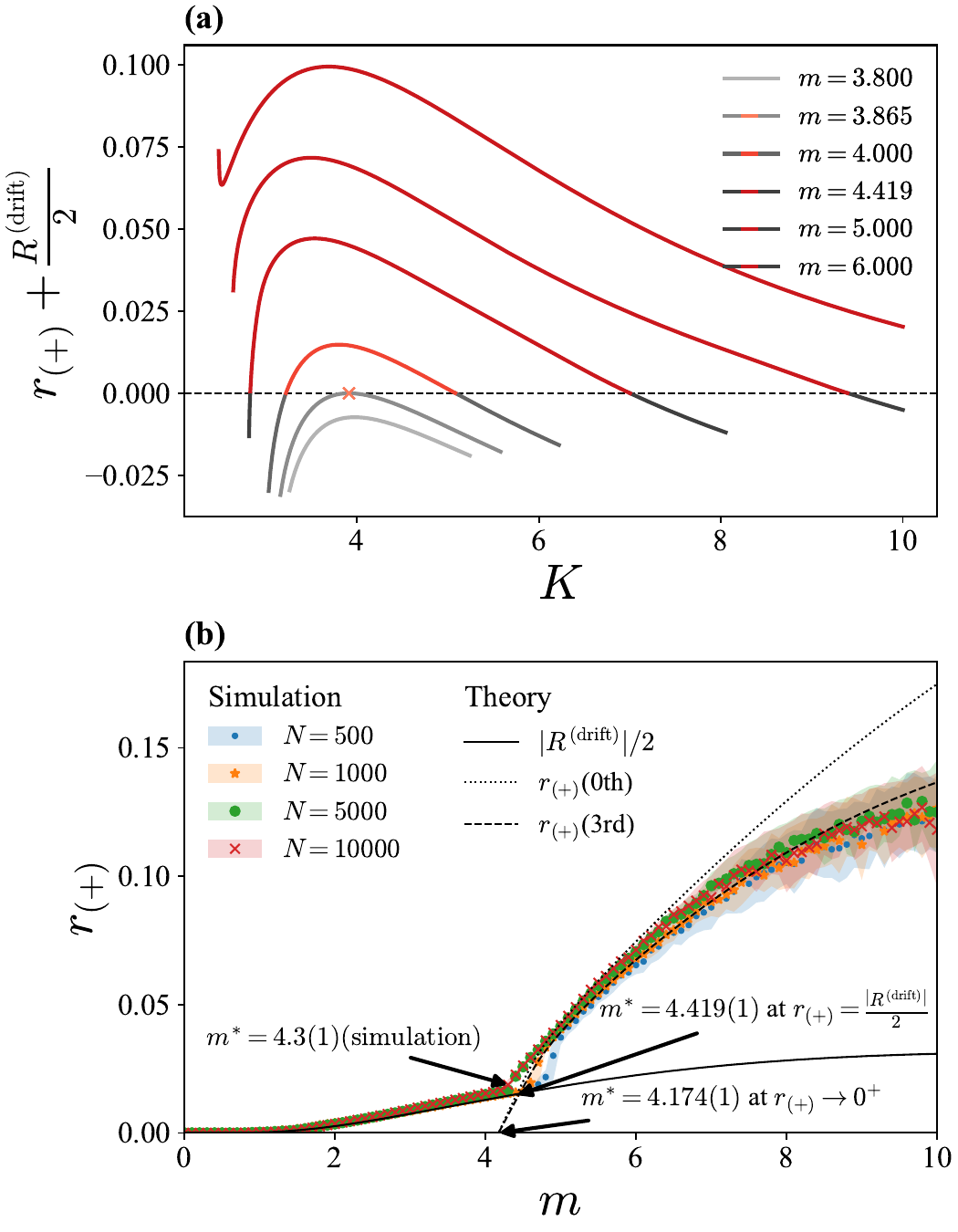}
    \caption{The crossover mass $m^{*}$ at which the secondary clusters can emerge or vanish. (a) The quantity $r_\p + {R^{\dr}}/{2}$ as a function of $K$ for various $m$ from $m = 3.8$ to $m = 6$. Secondary clusters can exist when $r_\p + {R^{\dr}}/{2} \ge 0$. The marginal value for the onset of the secondary clusters is $\tilde{m}^{*} \simeq 3.865$. This prediction does not reflect the prerequisite condition $R>0$, giving the infeasible regime as in Fig.~\ref{fig:sim_thr}(d). (b) The secondary synchrony $r_\p$ as a function of $m$ at $K = 7$ with $t = 10\,100$, $\Delta t = 50$, and $\mathrm{d}m = 0.1$. The zeroth-order and third-order approximation results are represented by dotted and dashed lines, respectively. The arrows indicate the crossover points: $m^{*} = 4.3(1)$ from simulation, $m^{*} = 4.174(1)$ from theory based on $r_\p \to 0^{+}$, and $m^{*} = 4.419(1)$ from theory based on $r_\p = |R^{\dr}/2|$ (our conjecture). The shaded regions represent the error bars computed over 100 realizations.
    }
    \label{fig:critical_m}
\end{figure}

In Fig.~\ref{fig:sim_thr}(d), the secondary synchronization may fail to emerge when the mass 
$m$ is either too small or too large, even at large $K$. In particular, secondary clusters never 
form when $m$ is too small, regardless of the coupling strength, whereas for sufficiently large 
$m$ they appear only within an intermediate range of $K$ and disappear again when $K$ becomes 
too large. These observations indicate the existence of a crossover mass $m^{*}$ and oscillatory regimes in which secondary synchronization can occur, and thus whether the global synchrony $R$ exhibits an oscillatory behavior.

Note that the final form of $r_\p$ in Eq.~(\ref{eq:rplus_selfcon}) can be written as a function of 
the mass $m$ for a given $K$. This allows us to determine the crossover mass $m^{*}$ by examining 
the condition $r_\p \to 0^{+}$. Under this condition, we obtain
\begin{equation}
    K r_\p m^{*} 
    \int_{\omega_\p^{*}+\delta\Omega + K r_\p b_S(a^{*})}^{\infty} 
        g(\Omega)\,\frac{1}{1+(m^{*})^{2}\Omega^{2}} \, \mathrm{d}\Omega \nonumber \\
    =  
    2 \int_{\omega_\p^{*}}^{\omega_\p^{*}+\delta\Omega + K r_\p b_S(a^{*})} 
        g(\Omega)\,
        \sqrt{1 - 
            \left(
                \frac{\Omega - \omega_\p^{*} - \delta\Omega}{K r_\p}
            \right)^{2}
        } 
        \, \mathrm{d}\Omega~, \nonumber \\
\end{equation}
where an asterisk ($^*$) indicates that the corresponding quantity explicitly depends on the 
crossover mass $m^{*}$.
We suggest another conjecture for the crossover mass $m^{*}$, motivated by the intriguing 
behavior of $\hat{r}_\p$ observed in the simulations, where 
$\hat{r}_\p = \max\left(r_\p, | {R^{\dr}}/{2} | \right)$ 
[see Fig.~\ref{fig:critical_m}(a)]. 
Assuming, as numerically supported, that $|Z^{\dr}| = 2 |z_\p|$ for the drifting oscillators, 
we obtain 
$|R^{\dr}| = |2 r_\p \langle \cos \phi_\p \rangle_t |$, 
which guarantees the inequality 
$|r_\p| \geq | {R^{\dr}}/{2} |$.

Since this inequality always holds, the onset of secondary synchronization should occur 
precisely when the two sides become equal. This leads to the condition
\begin{equation}
    |r_\p| = \left| \frac{R^{\dr}}{2} \right|.
    \label{eq:rp_Rdr_2}
\end{equation}
This equality motivates our conjecture that the crossover mass $m^{*}$ appears at the point 
where the secondary synchrony becomes comparable to the drifting contribution. In physical terms, this condition marks the point at which the coherent contribution of the secondary synchronized cluster becomes comparable to, and is no longer masked by, the background drift component, so that the secondary cluster can appear itself as a distinct collective structure. Note that condition $r_\p\to0^+$ gives a lower bound, whereas the condition in Eq.~(\ref{eq:rp_Rdr_2}) gives a sharper but conjectural bound. As will be discussed later, however, the condition $r_\p \to 0^{+}$ can also yield a more appropriate estimate of $m^{*}$. The phase boundaries obtained from both conditions, 
$r_\p \to 0^{+}$ and 
$|r_\p| = | {R^{\dr}}/{2} |$, 
are shown in Figs.~\ref{fig:sim_thr}(d) and~\ref{fig:sim_thr}(f).

Based on the formulation in Eq.~(\ref{eq:z_drift}), the global drift term can be written as $R^{\dr}= -2\int_{\omega_\p}^{\infty} g(\Omega) \frac{K R m}{2(1+m^2 \Omega^2)} \, \mathrm{d}\Omega$, 
which carries a minus sign (thus $|R^{\dr}| = -R^{\dr}$).  
Using this expression, we consider the combined quantity 
$r_\p + {R^{\dr}}/{2}$, which corresponds to the recast form of the equality condition proposed in the conjecture [Eq.~(\ref{eq:rp_Rdr_2})].  
This quantity is plotted as a function of $K$ for various values of $m$ in 
Fig.~\ref{fig:critical_m}(a).

The secondary clusters exist when $r_\p + {R^{\dr}}/{2} \geq 0$.
For small $m$, the value of $r_\p + {R^{\dr}}/{2}$ remains negative for all $K$, indicating
the absence of secondary clusters and thus the absence of oscillations in $R$.
For large $m$, the region satisfying $r_\p + {R^{\dr}}/{2} \geq 0$ widens but
remains confined within a finite range of $K$, corresponding to the oscillatory regime.
From this analysis, the marginal value of the crossover mass is identified as
$\tilde{m}^{*} \simeq 3.865$, corresponding to the first onset of secondary synchrony.
We emphasize that the condition $r_\p + {R^{\dr}}/{2} \geq 0$ does not determine the global 
synchrony threshold $K_c$.  
Secondary clusters can exist only when both 
$r_\p + {R^{\dr}}/{2} \geq 0$ and $K \geq K_c$ are satisfied, and the region with $r_\p + {R^{\dr}}/{2} \geq 0$ but $K < K_c$ (no global synchrony) is numerically infeasible and appears as the white regime inside the boundary in Fig.~\ref{fig:sim_thr}(d). 
For example, for $K=7$, one finds $m^{*}=4.419$ from the condition $r_\p + {R^{\dr}}/{2}=0$ in Fig.~\ref{fig:critical_m}(a). The corresponding secondary  synchrony $r_\p$ as a function of $m$ is shown in Fig.~\ref{fig:critical_m}(b), and one can see that the secondary clusters exist for any $m\gtrsim {m}^{*}$. The higher-order correction in $\delta\Omega$ discussed in Sec.~\ref{subsec:secondary_cluster_eq} improves the quantitative agreement, but the zeroth-order approximation in Eq.~(\ref{eq:theta_average}) is sufficient to capture the behavior near the transition point. As mentioned earlier, the simulation result $\hat{r}_\p$ follows the larger of the theoretical  values $r_\p$ and $|{R^{\dr}}/{2} |$.  We observe that the crossover mass $m^{*}$ predicted by our conjecture is closer to the numerically obtained value.

This analysis not only identifies the onset of the secondary clusters but also determines the point at which the secondary synchronization disappears, namely when $r_\p + {R^{\dr}}/{2} < 0$. In this sense, our criterion predicts the oscillatory regimes of the coupling strength $K$ in which the global synchrony $R$ exhibits oscillatory behavior, thereby providing a basis for the following analysis.

\section{Forward Process and Oscillatory Regimes of Synchrony}
\label{sec:forward}

\begin{figure*}
    \centering
    \includegraphics[width=0.5\linewidth]{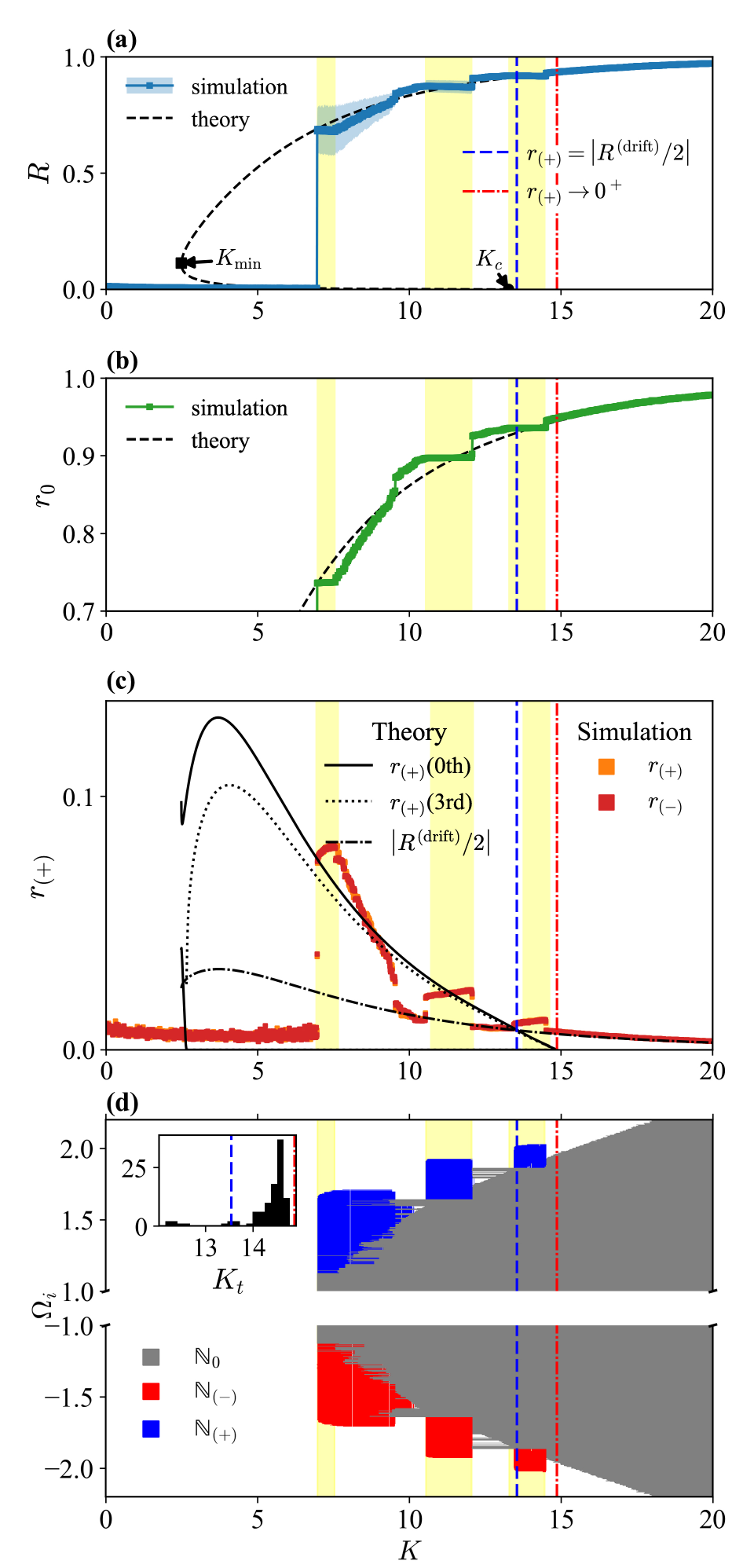}
    \caption{The analytic and simulation results of the forward process (increase in $K$, $\mathrm{d}K=0.01$) at $m=6$ and $N=5\,000$, evaluated at $t=2\,500$ with $\Delta t=50$. (a) The global synchrony $R$ versus the coupling strength $K$. The critical points $K_{\mathrm{min}}$ and $K_c$ are theoretically evaluated as the smallest $K$ for which $r_0>0$ and the value of $K$ at which $r_0 \to 0^{+}$, respectively. (b) The main-cluster synchrony $r_0$ versus $K$. (c) The secondary synchrony $r_\p$ versus $K$. The zeroth-order and third-order approximation results are represented by solid and dotted lines, respectively.
    (d) The intrinsic frequency $\Omega_i$ versus $K$ for all oscillators. White denotes drifting oscillators in $\mathbb{N}_{\Omega \gtrless \Omega_{(\pm)}}$, gray denotes the main cluster $\mathbb{N}_0$, and blue and red denote the secondary clusters $\mathbb{N}_{(\pm)}$. 
    The two dashed vertical lines indicate the theoretical boundaries $r_\p = \left|R^{\dr}/2\right|$ and $r_\p \to 0^{+}$, and the vertically shaded regions mark the plateau region of $R$. Inset: histogram of the terminal point $K_t$ of the oscillatory regime over 100 realizations.
}
    \label{fig:forward}
\end{figure*}

The system with inertia exhibits hysteresis phenomena~\cite{tanaka1997first}, in which the global order parameter $R$ shows different behaviors in the forward (increasing $K$) and backward (decreasing $K$) processes. The behavior of global synchrony has been well known and extensively analyzed~\cite{olmi2014hysteretic, gao2018self, gao2021synchronized}. Theoretically, the two processes are distinguished by the boundary value $\omega_\p$ used in the integration. However, the expression of $\omega_\p$ for the backward process in Eq.~(\ref{eq:lowerlimit}) corresponds to the branch starting from the fully synchronized state ($R=1$). We find that starting from $R<1$ in the backward process requires nontrivial modification of both the self-consistent equation and the integration boundary.
We therefore leave this issue for future work. Based on this reasoning, we focus on analyzing the forward process herein. In particular, we investigate its distinctive features through the secondary synchrony $r_\p$, building on our theoretical foundation beyond the global synchrony.

In the forward process, corresponding to increasing $K$ by $\mathrm{d}K$ and initializing the system from the state at $K-\mathrm{d}K$, the staircase-like increase in $R(K)$ appears only in numerical simulations for a single realization. This non-smooth behavior, illustrated in Fig.~\ref{fig:forward}(a), is closely related to the oscillatory dynamics of $R(t;K)$, as previously reported for $m=6$ in Ref.~\cite{olmi2014hysteretic}, where the staircase-like transition was demonstrated. It arises from the intermittent reorganization of oscillator clusters.
However, the range of $K$ over which this staircase pattern persists, that is, the region where $R(t;K)$ exhibits oscillatory behavior, has not been explicitly analyzed so far. Here, we aim to delineate this regime more clearly by employing the secondary-cluster analysis method.

We show the synchrony $r_0$ of the main cluster and $r_\p$ of the secondary clusters in Figs.~\ref{fig:forward}(b) and \ref{fig:forward}(c). Both $r_0$ and $r_\p$ display trends consistent with the theoretical expressions. When the secondary clusters exist, $r_\p$ closely follows the theoretical line, whereas when they vanish and only the main synchronized and drifting populations remain, $r_\p$ approaches $|R^{\dr}/2|$ as shown in Fig.~\ref{fig:forward}(c). As the system alternates between states with and without secondary clusters, the behavior of $R$ reflects these transitions, exhibiting plateaus and sudden increases that manifest as the staircase pattern in the forward process.

To illustrate how the oscillatory behavior of $R$ appears in the microscopic dynamics, we display the intrinsic frequencies of oscillators, with colors indicating their cluster index for each value of $K$, as shown in Fig.~\ref{fig:forward}(d). As $K$ increases, a significant fraction of oscillators abruptly form secondary clusters, and the corresponding plateaus in $R$ emerge, consistent with previous studies. 
The formation and disappearance of these secondary clusters are closely associated with the plateau–jump structure of $R$. When the secondary clusters exist [colored in red and blue in Fig.~\ref{fig:forward}(d)], $R$ remains nearly constant. When they vanish [red and blue regions absent in Fig.~\ref{fig:forward}(d)], the remaining oscillators are absorbed into the main cluster, producing a sudden increase in $R$. Such sudden increases in $R$ are reminiscent of explosive synchronization and, more broadly, of percolation-like collective growth.

This secondary-cluster analysis method provides a quantitative estimate of the oscillatory regimes of the staircase behavior of $R$, which has not been quantified in conventional theoretical approaches based solely on $R$. As discussed in Sec.~\ref{subsec:critical_mass}, secondary clusters exist when $r_\p > 0$ (a less strict boundary) or $r_\p > |R^{\dr}/2|$ (a stricter boundary suggested by the conjecture), provided that the main cluster exists ($K \geq K_c$). This allows us to identify the crossover value of $K$ at which the secondary clusters completely vanish, corresponding to the termination of the oscillatory behavior of $R$, that is, the disappearance of the staircase-like pattern observed in numerical simulations.

The termination point $K_t$ is predicted by either $r_\p \to 0^{+}$ or $r_\p = |R^{\dr}/2|$, and it varies from realization to realization. The histogram of $K_t$ is shown in the inset of Fig.~\ref{fig:forward}(d). The numerical crossover boundary also lies between $r_\p \to 0^{+}$ and $r_\p = |R^{\dr}/2|$, as seen in Fig.~\ref{fig:sim_thr}(d), and in most cases $K_t$ is closer to the condition $r_\p \to 0^{+}$. Although stochastic effects likely play a role in determining the realized $K_t$, the mechanism selecting which of the two boundaries becomes dominant remains unclear. Nevertheless, the oscillatory regimes, equivalent to the staircase pattern, can be effectively predicted by our secondary-cluster analysis method.

\section{Conclusion}
\label{sec:conclusions}
In this work, we have developed a theoretical and numerical framework to clarify how secondary synchronized clusters shape the oscillatory behavior of the global order parameter in the Kuramoto model with inertia. Building on a refined self-consistent description of the secondary synchrony $r_\p$ in Eq.~(\ref{eq:rplus_selfcon}), we have identified the crossover mass $\tilde{m}^*\simeq 3.865$ that determines whether secondary clusters can emerge at all [Fig.~\ref{fig:critical_m}]. The non-smooth, staircase-like behavior of $R(K)$ in the forward process is governed by the presence and disappearance of secondary clusters, consistent with previous findings~\cite{olmi2014hysteretic,kim2025cluster}. By combining either the theoretical boundary condition $r_\p \to 0^{+}$ or $r_\p = |R^{\dr}/2|$, we have quantified the oscillatory regimes of the global synchrony, providing a predictive criterion for the termination point $K_t$ [Figs.~\ref{fig:critical_m}~and~\ref{fig:forward}]. These results establish secondary-cluster dynamics as a fundamental mechanism underlying the oscillatory synchrony in the inertial oscillator system.

The relevance of secondary synchronization also extends to practical settings in which oscillatory behavior is undesirable. In power-grid models with inertial generators~\cite{filatrella2008analysis, dofler2013}, for example, persistent oscillations of the global frequency are often associated with reduced stability margins and an increased risk of transient failures. Since secondary clusters correspond to the regime in which the global order parameter $R$ enters a limit-cycle state rather than a stable fixed point, identifying the crossover mass $\tilde{m}^{*}$ and the oscillatory regimes in the $(K,m)$ parameter space provides a useful guideline for avoiding oscillatory responses. Our framework, therefore, offers insight into how coupling strength and inertia should be selected to suppress the emergence of secondary clusters, thereby promoting stable, fixed-point synchrony in inertia-dominated systems such as synchronous power grids.

This study still has several limitations. First, our analysis focuses exclusively on the emergence of secondary clusters and does not incorporate higher-order clustering phenomena. For sufficiently large $m$, additional clusters beyond the secondary ones may form, as also noted in recent studies~\cite{kim2025cluster,gao2021synchronized, park_hybrid_2024}, which can partially account for the discrepancies between zeroth-order theoretical predictions and numerical observations of $r_\p$ in Fig.~\ref{fig:sim_thr}. Second, our theoretical formulation has been developed only for the forward process. Extending the framework to the backward process would require a consistent treatment of partially synchronized initial states ($R<1$), which remains unresolved despite the structural simplicity of the system considered here, namely the complete graph. Furthermore, exploring the bistable regime ($b_S<b<b_L$) remains an intriguing and challenging topic. Third, we have considered a homogeneous mean-field setting, in which all oscillators are globally and uniformly coupled. Based on previous studies of dynamical processes (e.g., epidemic spreading) on complex networks, we expect that this mean-field-based secondary-cluster method can be extended to complex networks when the conditions of dense and uncorrelated or weakly correlated structures are satisfied (e.g., Erd\H{o}s--R\'{e}nyi random graphs and small-world networks with low clustering~\cite{RMP2015}). In real systems, power-grid networks are often regarded as relatively homogeneous structures with low clustering. In contrast, heterogeneous networks such as scale-free (SF) networks require a heterogeneous mean-field approach, i.e., a generalization of the present framework. While inertia-induced synchronization is expected to persist qualitatively in homogeneous networks, the situation is less straightforward in SF networks, where the subtle competition between many low-degree nodes and a few high-degree nodes can make the dominant dynamical contribution nontrivial. A recent study has shown that the population of small-degree nodes in SF networks can significantly affect macroscopic behavior such as network percolation~\cite{Kim2019pre}. These generalizations are non-trivial and fall outside the current scope, but they represent potential directions for future investigation. Finally, beyond these methodological limitations, there remain important dynamical aspects of inertial Kuramoto systems that are not addressed in the present work. In particular, the relaxation dynamics toward synchronized or oscillatory states, which play a central role in transient stability, warrant separate investigation~\cite{lee2014finite}. These limitations are not merely constraints of the present analysis but also point to natural directions for future work, where the secondary-cluster framework could be generalized and tested in broader classes of inertial synchronization systems. 

\section{Acknowledgments} \label{sec:acknowledgements}
This work was supported by the National Research Foundation (NRF) of Korea through Grant Numbers NRF-2023R1A2C1007523 (S.-W.S.), RS-2024-00341317 (M.J.L.) and a New Faculty Research Grant of Pusan National University, 2025 (M.J.L.). This work was also partly supported by Korea Research Institute for defense Technology  planning and advancement  (KRIT) - Grant  funded  by  Defense  Acquisition  Program Administration  (DAPA),  South  Korea  (KRIT-CT-23-026, Integrated  Underwater  Surveillance  Research  Center  for Adapting Future Technologies, 2023–2029). We thank APCTP, Pohang, Korea for their hospitality during the Topical Research Program [APCTP-2025-T04], from which this work greatly benefited.


\section*{Data availability}
Data and simulation codes are available from the corresponding authors upon reasonable request.








\printcredits


\bibliography{kuramoto}

\providecommand{\noopsort}[1]{}\providecommand{\singleletter}[1]{#1}%

\end{document}